\documentclass[letter]{aa}
\usepackage{amsmath}
\usepackage{xcolor}
\usepackage{txfonts}
\usepackage{natbib}
\bibpunct{(}{)}{;}{a}{}{,} 
\usepackage{xspace}
\usepackage{graphicx}
\usepackage{ifthen}
\usepackage{float}

\newcommand{\ddel}[2]{\frac{\partial #1}{\partial #2}}
\newcommand{\St}{\text{St}\xspace}
\newcommand{\rhos}{\ensuremath{\rho_\text{s}}\xspace}
\newcommand{\rhodust}{\ensuremath{\rho_\mathrm{d}}\xspace}
\newcommand{\rhogas}{\ensuremath{\rho_\mathrm{g}}\xspace}

\newcommand{\csound}{\ensuremath{c_\mathrm{s}}\xspace}

\newcommand{\Ok}{\ensuremath{\Omega_\mathrm{k}}\xspace}
\newcommand{\Vk}{\ensuremath{V_\mathrm{k}}\xspace}
\newcommand{\Siggas}{\ensuremath{\Sigma_\mathrm{g}}\xspace}
\newcommand{\Sigdust}{\ensuremath{\Sigma_\mathrm{d}}\xspace}

\newcommand{\alphat}{\ensuremath{\alpha_\text{t}}\xspace}
\newcommand{\alphamm}{\ensuremath{\alpha_\text{mm}}\xspace}
\newcommand{\vdr}{\ensuremath{u_{\mathrm{d},r}}\xspace}
\newcommand{\vdp}{\ensuremath{u_{\mathrm{d},\phi}}\xspace}
\newcommand{\vgr}{\ensuremath{u_{\mathrm{g},r}}\xspace}
\newcommand{\vgp}{\ensuremath{u_{\mathrm{g},\phi}}\xspace}
\newcommand{\arccot}{\ensuremath{\mathrm{arccot}}\xspace}

\makeatletter \if@referee
    \usepackage{endfloat}
    \newcommand{\highlight}[1]{\textbf{\textcolor{red}{#1}}}
\else
    \newcommand{\highlight}[1]{#1}
\fi \makeatother

\usepackage[author={Til Birnstiel}]{pdfcomment} 

\hypersetup{
  pdfstartview=FitH,
  linkcolor=blue,
  anchorcolor=blue,
  citecolor=blue,
  urlcolor=blue,
  colorlinks=true,
  breaklinks}

\begin{document}
\title{Lopsided dust rings in transition disks}
\titlerunning{Lopsided dust rings in transition disks}
\author{T. Birnstiel\inst{1}\fnmsep\inst{2} \and C.P. Dullemond\inst{3} \and P.
Pinilla\inst{3}}
\authorrunning{T.~Birnstiel, C.P. Dullemond, P. Pinilla}
\institute{
    Excellence Cluster Universe, Technische Universit\"at M\"unchen, Boltzmannstr. 2, 85748 Garching, Germany \and
    Harvard-Smithsonian Center for Astrophysics, 60 Garden Street, Cambridge, MA 02138, USA\and
    Heidelberg University, Center for Astronomy (ZAH), Institute for Theoretical Astrophysics, Albert Ueberle Str. 2, 69120 Heidelberg, Germany
}
\date{\today}

\abstract
{Particle trapping in local or global pressure maxima in protoplanetary disks is one of the new
paradigms in the theory of the first stages of planet formation. However, finding observational
evidence for this effect is not easy. Recent work suggests that the large
ring-shaped outer disks observed in transition disk sources may in fact be lopsided and constitute
large banana-shaped vortices.}
{We wish to investigate how effective dust can accumulate along the azimuthal direction. We also
want to find out if the size-sorting resulting from this can produce a detectable signatures
at millimeter wavelengths.}
{
To keep the numerical cost under control we develop a 1+1D method in which the azimuthal variations
are treated separately from the radial ones.
The azimuthal structure is calculated analytically for a steady-state between mixing and azimuthal
drift. We derive equilibration time scales and compare the analytical solutions to time-dependent
numerical simulations.}
{We find that weak, but long-lived azimuthal density gradients in the gas can induce
very strong azimuthal accumulations of dust. The strength of the accumulations depends on the
P\'eclet number, which is the relative importance of advection and diffusion.
We apply our model to transition disks and our simulated observations show that this
effect would be easily observable with ALMA and in principle allows to put constraints on the
strength of turbulence and the local gas density.}
{}

\keywords{accretion, accretion disks -- protoplanetary disks -- stars: pre-main-sequence,
circumstellar matter -- planets and satellites: formation -- submillimeter: planetary systems}

\maketitle

\section{Introduction}\label{sec:introduction}
The process of planet formation is thought to start with the growth of dust aggregates in a
protoplanetary disk through coagulation \citep[see e.g., the review by ][]{Blum:2008p1920}. The idea
is that these aggregates get successively bigger until either gravoturbulent processes set in
\citep{Goldreich:1973p11184,Johansen:2007p4788,Cuzzi:2008p12253} or planetesimals are formed
directly via coagulation \citep{Weidenschilling:1977p865,Okuzumi:2012p18448,Windmark:2012p18492}.
One of the main unsolved problems in these scenarios is the problem of excessive radial drift
\citep{Weidenschilling:1977p865,Nakagawa:1986p2048,Brauer:2007p232}. The origin of this problem lies
in the sub-Keplerian motion of the gas in the disk, caused by the inward pointing pressure gradient.
The dust particles in the disk, however, feel only the friction with the sub-Keplerian gas and as
they grow to larger sizes, the reduced surface-to-mass ratio causes them to fall toward the star
with speeds up to 50~m~s$^{-1}$ \citep{Whipple:1972p4621,Weidenschilling:1977p865}. Thus, if a
particle grows, it will sooner or later get ``flushed'' toward the star before it can grow very
large \citep{Brauer:2008p212,Birnstiel:2010p9709,Birnstiel:2012p17135}.
This is what we call the \emph{radial drift barrier}, and this still poses one of the main unsolved
problems of the early phases of planet formation.

A possible solution to this problem might lie in the concept of ``particle traps''. This idea has
been proposed already some time ago in the context of anticyclonic vortices by
\citet{Barge:1995p11993} and \citet{Klahr:1997p19300}, as well as in/around turbulent eddies
\citep{Johansen:2005p8425,Cuzzi:2003p12258}. These vortex or eddy-related local pressure maxima act
as particle traps because particles tend to drift always towards higher pressure.
In the case of vortices and eddies they act on small scales. Particle traps on a global disk scale
have been proposed as well
\citep{Whipple:1972p4621,Rice:2006p19467,Alexander:2007p131,Garaud:2007p405,Kretke:2007p697,Dzyurkevich:2010p11360}
and shown to be conducive to planet formation \citep*{Brauer:2008p212}. Intermediate scale particle
traps may also occur from magnetorotationally driven turbulence: the so-called
zonal flows \citep{Johansen:2009p7441}.

If we want to observationally test whether this scenario of particle trapping actually occurs in
nature, we are faced with a problem. The Earth-forming region around a pre-main sequence star is
usually too small on the sky to be spatially resolved sufficiently well to test this trapping
scenario. Moreover, the optical depth of this inner disk region is likely to be too large to be able
to probe the mid-plane region of the disk. Fortunately, what constitutes the ``meter size drift
barrier'' at 1~AU is a ``cm size drift barrier'' at $\sim$~50~AU. Those disk regions are optically
thin at millimeter (mm) wavelength and particles in the mm size range can be spectroscopically
identified by studying the mm spectral slope
\citep{Testi:2001p9427,Natta:2004p3169,Ricci:2010p17766,Ricci:2010p9423}.
So the goal that has been pursued recently is to identify observational signatures of dust particle
trapping of millimeter-sized particles in the outer regions of disks, as a proxy of what happens in the
unobservable inner regions of the disk \citep{Pinilla:2012p16999,Pinilla:2012p18741}. The
\citet{Pinilla:2012p18741} paper suggests that the huge mm continuum rings observed in most of the
transition disks (\citealp{Pietu:2006p17014}, \citealp{Brown:2008p8893}, \citealp{Hughes:2009p17047}, 
\citealp{Isella:2010p17527}, \citealp{Andrews:2011p16142}), may in fact be large global particle
traps caused by the pressure bump resulting from, for example, a massive planet opening up a gap.

We have focused in these papers on the intermediate scale (zonal-flow-type) and the the large scale
(global) pressure bumps, simply because current capabilities of mm observatories (including
ALMA) are not yet able to resolve small scale structures such as vortices. The vortex trapping
scenario thus appears to remain observationally out of reach. However, a closer look at the
mm maps of transition disks suggest that some of them may exhibit a deviation from axial
symmetry.  For instance, the observations presented in \citet{Mayama:2012p19488} or the mm
images of \citet{Brown:2009p8895} suggests a banana shaped lopsided ring instead of a circular ring.
\citet{Regaly:2012p18108} proposed that these banana-shaped rings are in fact a natural consequence
of mass piled up at some obstacle in the disk.  Once the resulting ring becomes massive enough, it
becomes Rossby-unstable and a large banana-shaped vortex is formed that periodically fades and
re-forms with a maximum azimuthal gas density contrast of a few. \citet{Regaly:2012p18108} showed
that this naturally leads to lopsided rings seen in mm wavelength maps \citep[see also
earlier work by][]{Wolf:2002p19305} on radiative transfer predictions of observability of vortices
with ALMA). The formation of such Rossby-wave induced vortices was demonstrated before by
\citet{Lyra:2009p4812}, who showed that they may in fact (when they are situated much further
inward, in the planet forming region) lead to the rapid formation of planetary embryos of Mars mass.
\citet{Sandor:2011p19307} subsequently showed that this scenario may rapidly produce a 10~Earth mass
planetary core.

The goal of the current letter is to combine the scenario of forming a lopsided gas ring
\citep[e.g.,][]{Regaly:2012p18108} with the scenario of particle trapping and growth presented by
\citet{Pinilla:2012p18741}.
In Section~\ref{sec:analytical}, we will outline the physical effects involved and derive analytical
solutions to the dust distribution along the non-axisymmetric pressure bump and
Section~\ref{sec:sim_obs}, we will test the observability of these structures in resolved
\mbox{(sub-)mm} imaging and in the mm spectral index. Our findings will be summarized in
Sect.~\ref{sec:summary}.

\section{Analytical model}\label{sec:analytical}
Dust particles embedded in a gaseous disk feel drag forces if they move relatively to the gas. The
radial and azimuthal equations of motion have been solved for example by
\cite{Weidenschilling:1977p865} or \cite{Nakagawa:1986p2048} for the case of a axisymmetric, laminar
disk. It was found that particles drift inward towards higher pressure. In this paper, we will focus
on the case where a non-axisymmetric structure has formed a long lived, non-axisymmetric pressure
maximum in the disk. This pressure maximum is able to trap inward-spiralling dust particles. Like in
the aforementioned works we can solve for a stationary drift velocity, but unlike in
\cite{Nakagawa:1986p2048}, the \emph{radial} pressure gradient is zero at the pressure maximum while
the \emph{azimuthal} pressure gradient can be different from zero. This leads again to a systematic
drift motion of the dust particles towards the pressure maximum, but now in azimuthal instead of in
radial direction.

The equations of motion in polar coordinates relative to the Keplerian motion become
\citep[see][]{Nakagawa:1986p2048}
\begin{align}
\ddel{\vdr}{t} &= - A\, \rhogas  (\vdr -  \vgr) + 2 \Ok \vgp\\
\ddel{\vdp}{t} &= - A\, \rhogas  (\vdp -  \vgp) - \frac{1}{2} \Ok \vdr\\
\ddel{\vgr}{t} &= - A\, \rhodust (\vgr -  \vdr) + 2 \Ok \vgp
\end{align}
\begin{align}
\ddel{\vgp}{t} &= - A\, \rhodust (\vgp -  \vdp) - \frac{1}{2} \Ok \vgr - \frac{1}{r\,\rhogas} \ddel{P}{\phi},
\end{align}
where \vdr and \vdp are the $r$ and $\phi$ components of the dust velocity, respectively, \vgr and
\vgp the $r$ and $\phi$ components of the gas, \Ok the Keperian frequency, $P$ the gas pressure
and \rhodust and \rhogas the dust and gas densities. $A$ denotes the drag coefficient
\citep[see,][equations 2.3 and 2.4]{Nakagawa:1986p2048}. Solving above equations for the velocity along the
$\phi$ direction at the mid-plane ($z=0$) gives
\begin{equation}
\vdp = \frac{1}{\St+ \St^{-1}\left(1+X\right)^2}\, \frac{1}{\rhogas\,\Vk} \ddel{P}{\phi},
\label{eq:u_phi}
\end{equation}
where $X = \rhodust/\rhogas$ is the dust-to-gas ratio, \Vk the Keplerian velocity, and the Stokes
number\footnote{\St=1 typically corresponds to particles of mm to cm sizes in the outer disk. For
typical disk conditions $a \simeq 0.4\,\text{cm}\cdot \St \cdot
\Siggas/1\,\mathrm{g}\,\mathrm{cm}^{-2}$.} is given by
\begin{equation}
\St = \frac{\rhos\,a}{\rhogas\,\csound\,\sqrt{8/\pi}}\,\Ok,
\label{eq:St}
\end{equation}
with \rhos as internal density of the dust, particle radius $a$, and the isothermal sound speed
$\csound$.

\begin{figure}[tb]
    \centering
    \makeatletter \if@referee
        \resizebox{\hsize}{!}{\includegraphics{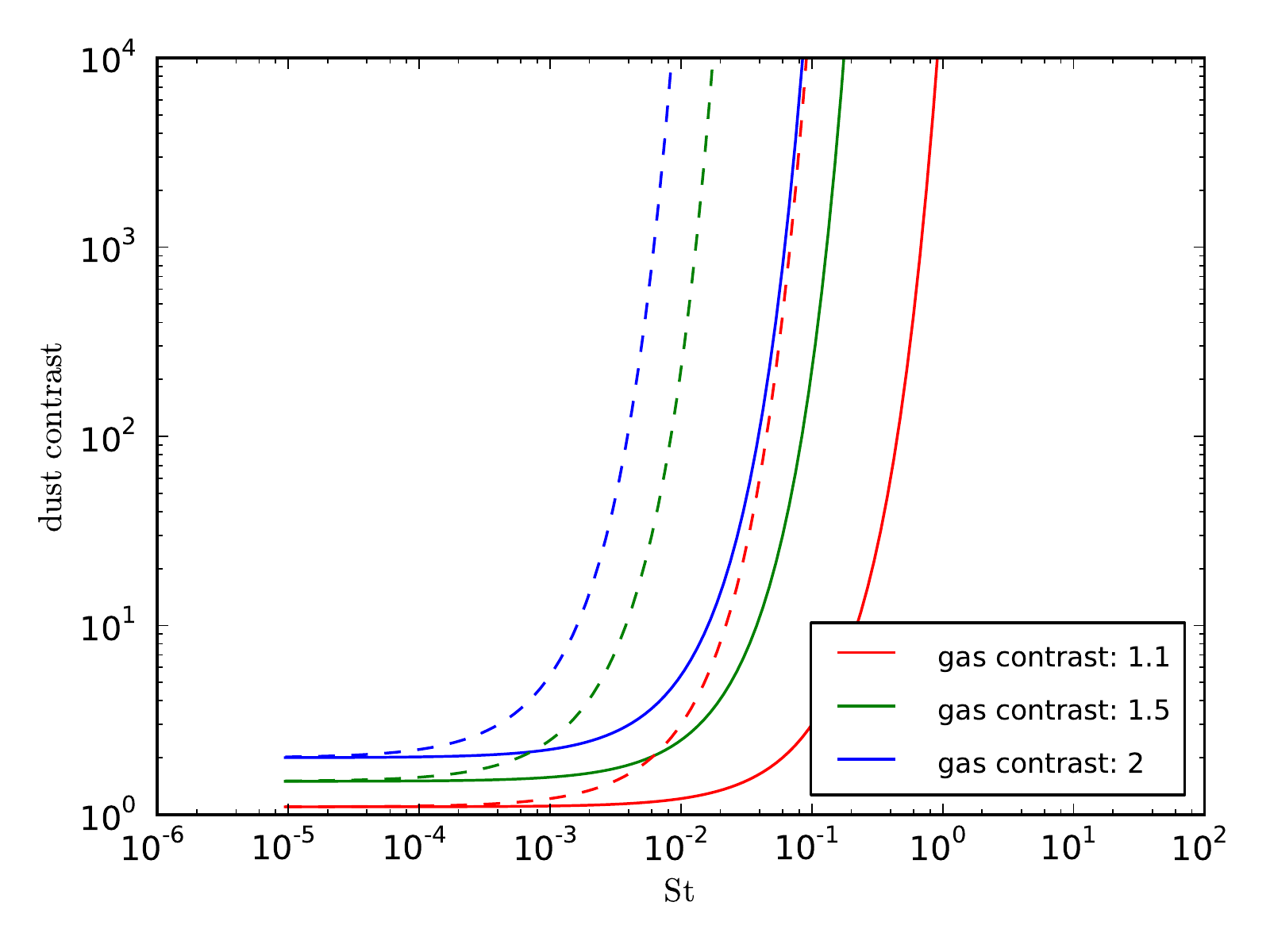}}
    \else
        \resizebox{\hsize}{!}{\includegraphics{plots/plot_contrast_analytical.pdf}}
    \fi \makeatother
  \caption{Contrast between the dust density in the azimuthal maximum and its surroundings for
  different gas density contrasts ($\rhogas^\mathrm{max}/\rhogas^\mathrm{min}$), Stokes
  numbers \St, as derived in Eq.~\ref{eq:enhancement}. Turbulence parameters are $\alphat=10^{-2}$
  (solid), and $\alphat=10^{-3}$ (dashed).}
  \label{fig:enhancement}
\end{figure}

Thus, dust is advected with the velocity given in Eq.~\ref{eq:u_phi} but it is also turbulently
stirred. Together, the evolution of the dust density along the ring is then described by
\begin{equation}
\ddel{\rhodust}{t} = \ddel{\,}{y}\left(\rhodust\,\vdp\right) -
\ddel{\,}{y}\left(D\,\rhogas\,\ddel{\,}{y}\left(\frac{\rhodust}{\rhogas}\right)\right),
\label{eq:adv_diff}
\end{equation}
where $y = r\,\phi$ is the coordinate along the ring circumference. We use a dust diffusion
coefficient $D$ according to \citet{Youdin:2007p2021},
$
D = D_\mathrm{gas}/(1+\St^2), 
$
where we assume the gas diffusivity to be equals to the gas viscosity, taken to be $\nu =
\alphat\,\csound^2/\Ok$, with \alphat as the turbulence parameter \citep[see][]{Shakura:1973p4854}.
Equation~\ref{eq:adv_diff} can be integrated forward in time numerically, but assuming that
the turbulent mixing and the drift term have reached an equilibrium and also assuming a low
dust-to-gas ratio, we can analytically solve for the dust density in a steady-state between mixing
and drifting, which yields
\begin{equation}
\rhodust(y) = C \cdot \rhogas(y) \cdot \exp\left[ -\frac{\St(y)}{\alphat}\right],
\label{eq:rhodust_analytical}
\end{equation}
where $C$ is a normalization constant and $\St(y)$ is the Stokes number which depends on $y$ via the
changes in gas density. Eq.~\ref{eq:rhodust_analytical} thus predicts the distribution of dust for
any given profile of the gas density \rhogas. The contrast between the position of the azimuthal
pressure maximum and its surrounding then gives
\begin{equation}
 \frac{\rhodust^\mathrm{max}}{\rhodust^\mathrm{min}}
=\frac{\rhogas ^\mathrm{max}}{\rhogas ^\mathrm{min}} \,
\exp\left[\frac{\St^\mathrm{min}-\St^\mathrm{max}}{\alphat}\right],
\label{eq:enhancement}
\end{equation}
which is plotted in Fig.~\ref{fig:enhancement}. $\St^\mathrm{max}$ and $\St^\mathrm{min}$ are the
Stokes numbers at the pressure maximum and minimum, respectively (note:
$\St^\mathrm{max} < \St^\mathrm{min}$). It can be seen that once the particles Stokes number becomes
larger than the turbulence parameter \alphat, the dust concentration becomes much stronger than the
gas concentration.
However, the time scales on which these concentrations are reached can be significant. In order to
get an estimate for this time scale, we compare the advection time scale $t_\mathrm{adv} = L/u$ and
the diffusion time scale $t_\mathrm{diff}=L^2/D$, with velocity $u$ and length scale $L$. The ratio
of these time scales, known as the P\'eclet number $\mathrm{Pe}$, which, using Eq.~\ref{eq:u_phi},
can be written as
\begin{equation}
\mathrm{Pe} 
    = \frac{t_\mathrm{diff}}{t_\mathrm{adv}}
    = \frac{\St}{\alphat}\,\frac{L}{\rhogas}\ddel{\rhogas}{y}
    \simeq \frac{\St}{\alphat}\frac{\Delta\rhogas}{\rhogas}.
\label{eq:peclet}
\end{equation}
It describes the relative importance of advection and diffusion and confirms that dust accumulations
occur only for particles with $\St\gtrsim\alphat$, because otherwise, diffusion dominates over advection
which means that variations in the dust-to-gas ratio are being smeared out. It also shows that for
those large particles the advection time scale is the shorter one, thus setting the time scale of
the concentration process, which can be written as

\begin{equation}
t_\mathrm{adv} =  \frac{\pi^2}{\delta\St} \left(\frac{H}{r}\right)^{-2} \frac{1}{\Ok},
\label{eq:t_adv}
\end{equation}
where $H = \csound/\Ok$ is the pressure scale height and we have used a mean velocity $\overline{u}
= \frac{1}{\pi} \left| \int_0^\pi u\, \mathrm{d}\phi\right| = \frac{\csound^2}{\pi \, \Vk}
\delta\St$, with
\begin{equation}
\delta \St = \left| \arccot\left(\St^\mathrm{min}\right) - \arccot\left(\St^\mathrm{max}\right) \right|,
\end{equation}
which for $\St<1$ simplifies to $\delta\St = \St^\mathrm{min}-\St^\mathrm{max}$.
Further, we define the pressure maximum and minimum to be at $\phi = 0$ and $\phi =
\pi$, respectively. As an example, at 35~AU, for $H/r=0.07$, a Stokes number of 0.2 and a gas
density contrast of $\rhogas^\mathrm{max}/\rhogas^\mathrm{min}=2$, the time scale is $3\times
10^5$~years, but could be as short as $10^2$ orbits for optimal conditions.
Any gas structure therefore has to be long-lived to cause strong asymmetries in the dust, making
e.g. asymmetries caused by a planet or long lived vortices the best candidates
\citep{Meheut:2012p19514}. If such an accumulation is formed and the gas asymmetry disappears, it
still takes $t_\mathrm{diff}$ to ``remove'' it, which at 35~AU is of the order of Myrs. It remains
to be shown whether short-lived, but reoccurring structures like zonal flows are able to induce
strong dust accumulations.

\begin{figure}[tb]
    \centering
    \makeatletter \if@referee
        \resizebox{\hsize}{!}{\includegraphics{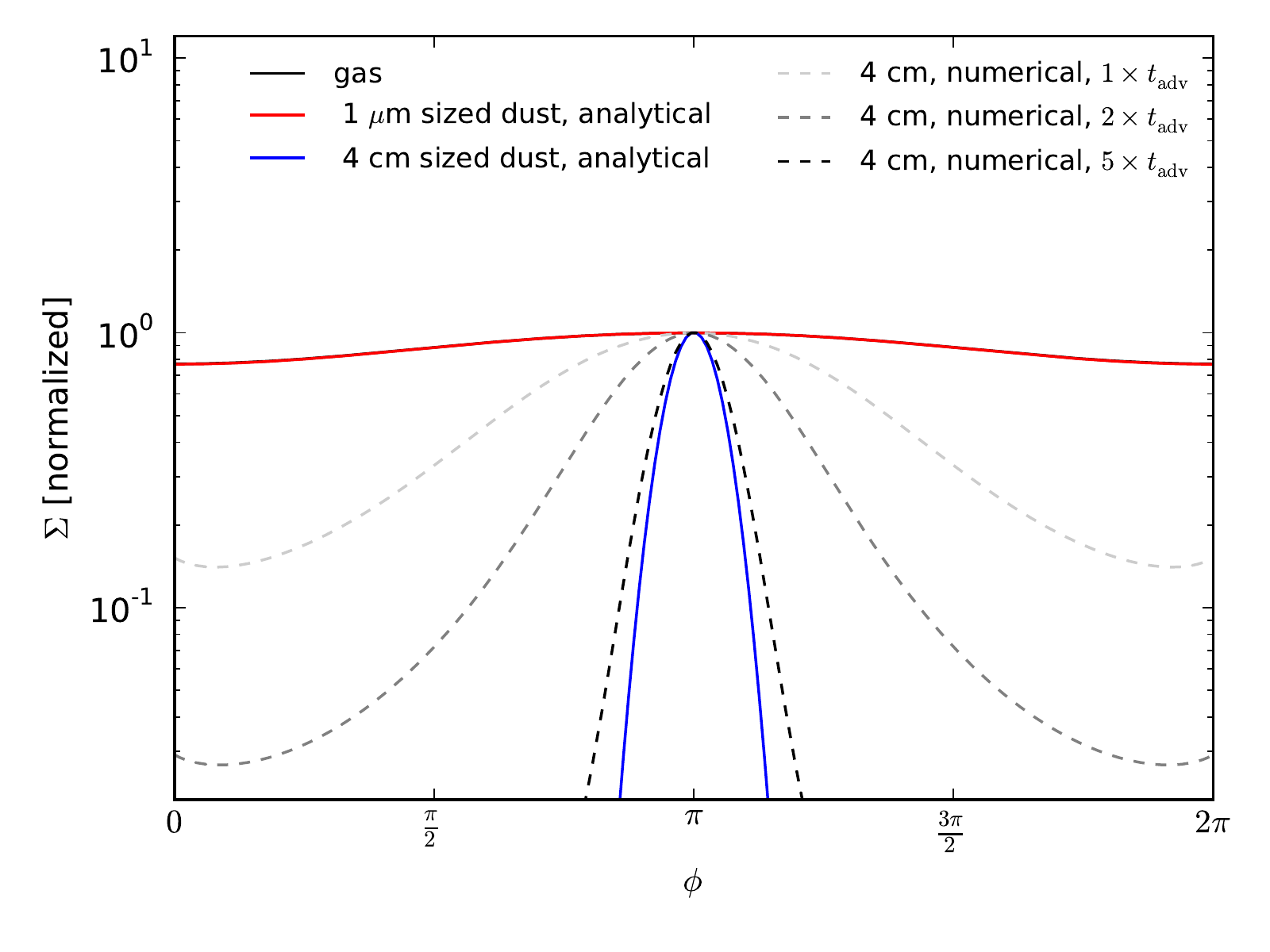}}
    \else
        \resizebox{\hsize}{!}{\includegraphics{plots/sinus_phi_profile.pdf}}
    \fi \makeatother
  \caption{Azimuthal steady state solutions for small (red line) and large dust grains (black line)
  for a given sinusoidal gas profile (black line, identical with red line). Dashed lines show numerical
  solutions at 1,2, and 5 advection time scales.}
  \label{fig:oneD_profiles}
\end{figure}

\section{Simulated Observations}\label{sec:sim_obs}
In the following, we will evaluate whether the dust structures we expect would be observable in
resolved mm images of ALMA. As radial gas surface density profile $\overline{\Sigma}_\mathrm{g}(r)$
\highlight{and temperature $T(r)$} we use the results presented in \citet{Pinilla:2012p18741}.
These simulations represent a disk of mass $M_\mathrm{disk} = 0.05\,M_\odot$ around a solar mass
star with a 15 Jupiter-mass planet at 20~AU. For simplicity the two-dimensional gas surface density
is then taken to be
\begin{align}
\Siggas(r,\phi) &= \overline{\Sigma}_\mathrm{g}(r) \cdot \left[1 + A(r) \cdot
                   \sin\left(\phi - \frac{\pi}{2}\right)\right]\\
A(r)            &= \frac{c-1}{c+1}\cdot \exp\left[-\frac{\left(r-R_\mathrm{s}\right)^2}{2\,H^2}\right]
\label{eq:gas_profile}
\end{align}
where $c = \Sigma_\mathrm{g,max}/\Sigma_\mathrm{g,min}$ is the largest contrast of the gas surface
density, taken to be 1.5. $R_\mathrm{s}$ is the position of the radial pressure bump. The dust size
distribution $\Sigdust(r,a)$ was also taken from the simulations of \citet{Pinilla:2012p18741}, and
distributed azimuthally using the analytical solution from Eq.~\ref{eq:rhodust_analytical} \highlight{(see
Fig.~\ref{fig:oneD_profiles}). We also confirmed the analytical solution and time scales by solving
Eq.~\ref{eq:adv_diff} numerically, as shown in Fig.~\ref{fig:oneD_profiles}.} This analytical solution
strictly only holds at the position of the radial pressure bump, but since most of the mm emission
comes from the large grains which in the simulations of \citet{Pinilla:2012p18741} are trapped near
the radial pressure maximum, this should be a reasonable approximation. Full two-dimensional
simulations will be needed to confirm this and to investigate the effects of shear.

\begin{figure}[tb]
    \centering
    \makeatletter \if@referee
        \resizebox{\hsize}{!}{\includegraphics{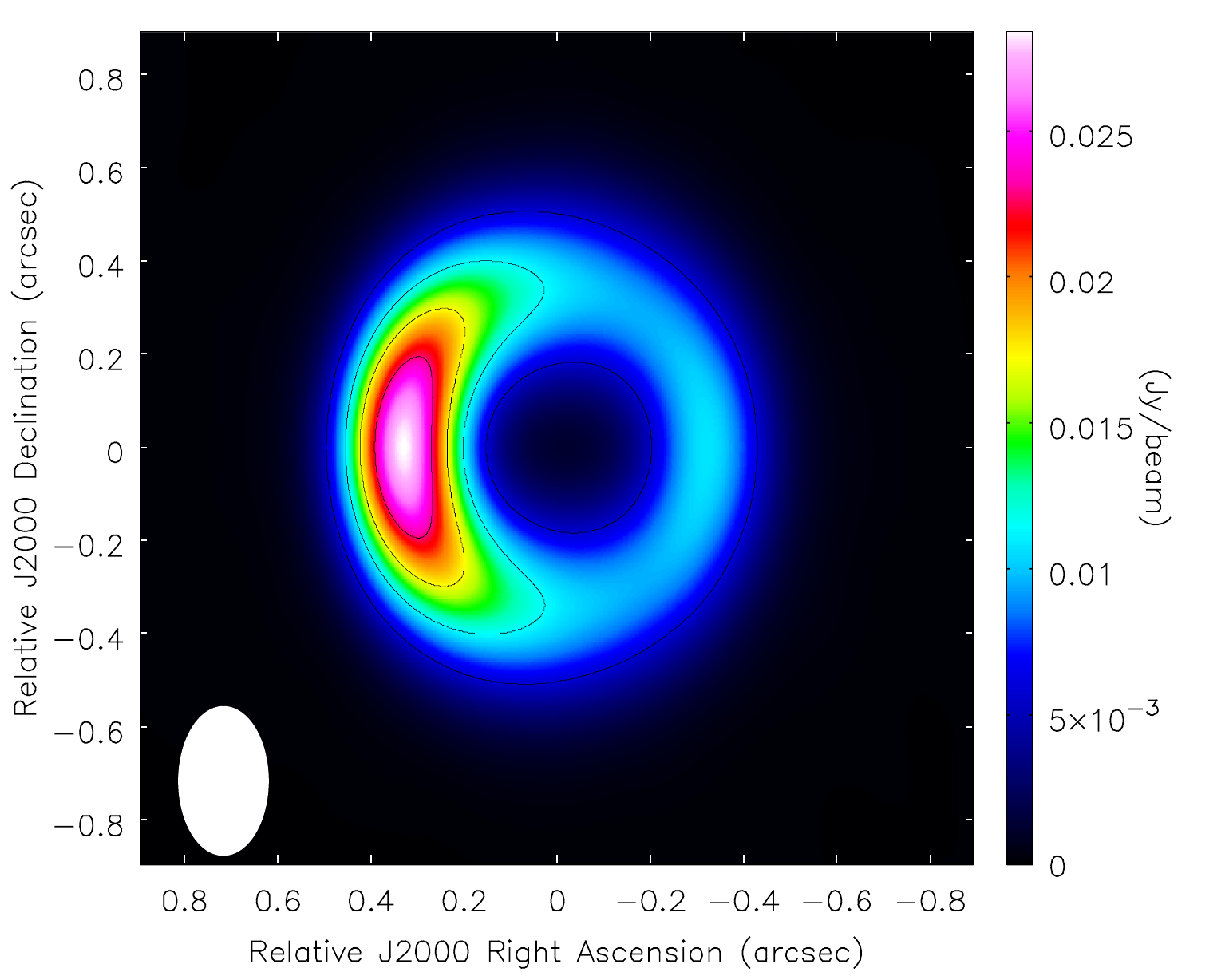}}
    \else
        \resizebox{!}{0.8\hsize}{\includegraphics{plots/banana_cycle1_band7_modified}}
    \fi \makeatother
  \caption{ALMA simulated images at 345~GHz with an observation time of
  2~hours. The total flux of the source is 0.13~Jy and the contour lines are at
  $\{2,4,6,8\}$ the rms ($\sigma~=~0.22~mJy$).
  }
  \label{fig:ALMA}
\end{figure}

To compare directly with current ALMA observations, we calculate the opacities for each grain size
at different wavelengths and assume spherical silicate grains  with optical constants for
magnesium-iron grains from the Jena
database\footnote{$\textrm{http://www.astro.uni-jena.de/Laboratory/Database/databases.html}$}.
The continuum intensity maps are calculated assuming that in the sub-mm regime, most of the disk
mass is concentrated in the optically thin region. We assume the same stellar parameters as in
\citet{Pinilla:2012p18741}, \highlight{azimuthally constant temperature $T(r)$}, typical source
distances (d~=~140~pc) and zero disk inclination. We run ALMA simulations using CASA (v. 3.4.0) at
345~GHz (band 7) and 675~GHz (band 9), shown in Fig.~\ref{fig:ALMA}. We consider 2~hours of
observation, the most extended configuration that is currently available with Cycle~1, generic
values for thermal and atmospheric noises and a bandwidth of $\Delta\nu~=~$7.5~GHz for continuum. At
these two different frequencies, it is possible to detect and resolve regions where the dust is
trapped creating a strong azimuthal intensity variation.

\begin{figure}[tb]
    \centering
    \makeatletter \if@referee
        \resizebox{\hsize}{!}{\includegraphics{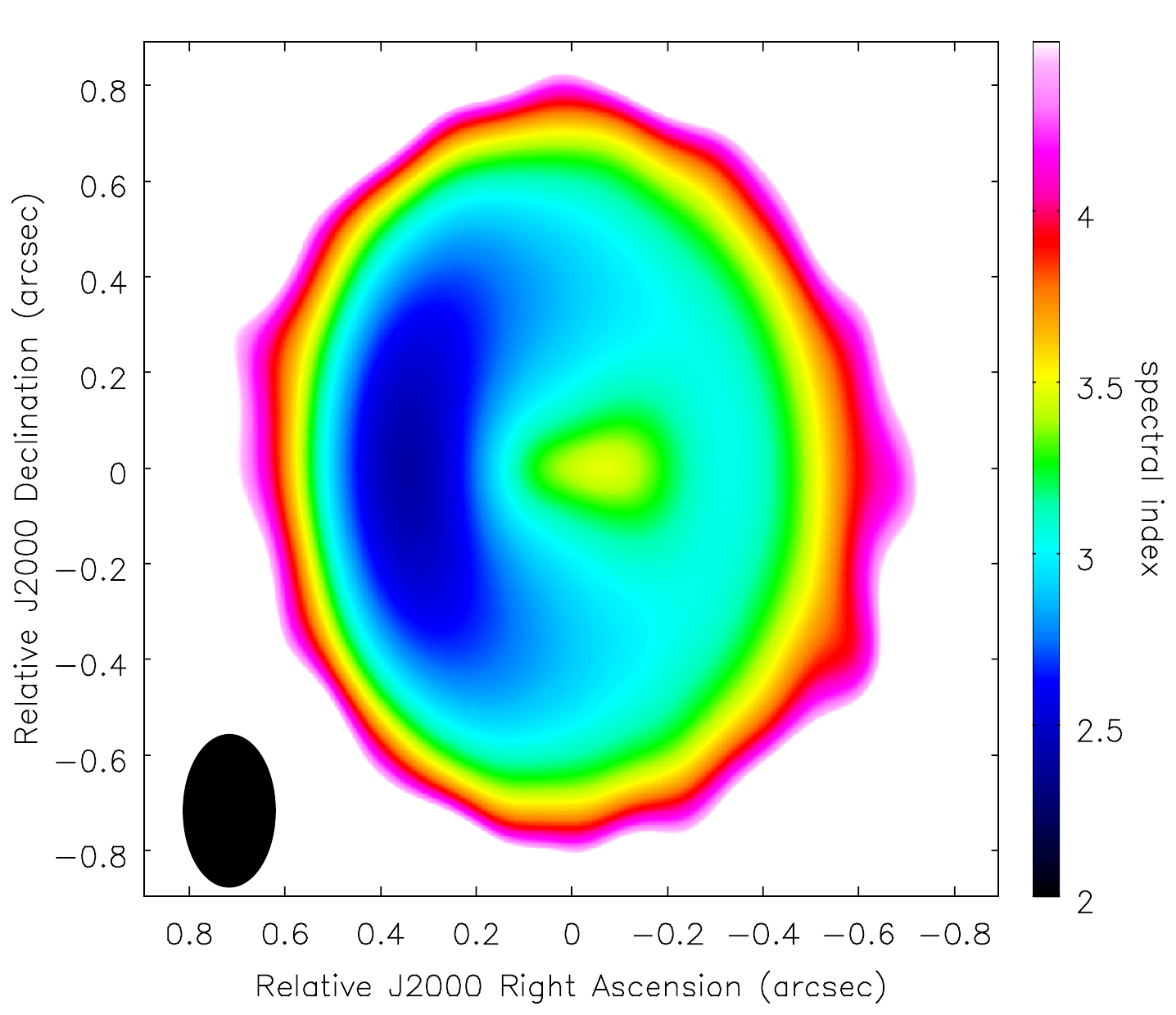}}
    \else
        \resizebox{!}{0.8\hsize}{\includegraphics{plots/spectral_index_band7_band9_modified}}
    \fi \makeatother
  \caption{Spectral index map \alphamm using simulated images at band~7 and 9.
  The antenna configuration is chosen such that the angular resolution is similar for both bands
  $\sim$~0.16'' ($\sim$~22AU at 140~pc).
  }
  \label{fig:ALMA_alpha_map}
\end{figure}

The spectral slope \alphamm of the spectral energy distribution $F_{\nu}~\propto~\nu^{\alpha}$ is
directly related to the dust opacity index at these long wavelengths
\citep[e.g.][]{Testi:2003p3390}, and it is interpreted in terms of the grain size
($\alphamm~\lesssim~3$ implies mm sized grains). With the simulated images of Fig.~\ref{fig:ALMA},
we compute the \alphamm map (Fig.~\ref{fig:ALMA_alpha_map}), considering an antenna configuration
that provides a similar resolution of  $\sim$~0.16'' ($\sim$~22AU at 140~pc) for each band.  This
resolution is enough to detect \alphamm variations along the azimuth, confirming that those are
regions where dust accumulates and grows due to the presence of an azimuthal pressure bump.

\section{Summary and conclusions}\label{sec:summary}
We have shown that weak, but long-lived azimuthal asymmetries in the gas density can cause very
efficient accumulation of dust at the position of the azimuthal pressure maximum. We have derived
analytical steady-state solutions for the dust distribution for any given azimuthal gas density
distribution and the time scales on which these distributions develop. Good agreement has been found
between the solutions and numerical simulations.

For this dust concentration mechanism to work, particles must have grown to larger sizes
($\St>\alphat$) such that the azimuthal drift becomes stronger than the turbulent diffusion. This
typically corresponds to particles of sub-mm to cm sizes. The strong concentration of the largest
grains leads to a size-sorting which is observable via low spectral indices at mm wavelengths and
also in lopsided banana shaped structures in resolved (sub-)mm images. Finding the size range where
the bifurcation between concentration and diffusion happens would put constraints on the turbulence
strength and the local gas density of the disk.

\begin{acknowledgements} 
We thank Ewine van Dishoeck, Simon Bruderer, Nienke van der Marel, Geoffrey Mathews, Hui
Li and the referee for useful discussions.
\end{acknowledgements}

\bibliographystyle{aa}
\bibliography{/Users/til/Documents/Papers/bibliography}

\makeatletter
\if@referee
\processdelayedfloats
\pagestyle{plain}
\fi
\makeatother
\end{document}